\documentclass[thmsa,10pt,a4paper,twoside]{article}
\usepackage{sw20lart}

\input tcilatex
\QQQ{Language}{
American English
}

\begin{document}

\author{MARIA CRISTINA NEACSU}
\title{A MAGNETIC\ MASS\ MODEL\ WITHIN\ THE\\
GENERAL\ RELATIVITY FRAMEWORK}
\date{Department of Physics, Technical University ''Gh. Asachi'' , Bd. Copou no
22, Iasi 6600, Romania}
\maketitle

\begin{abstract}
In this paper is presented a general-relativistic approach of cuasi-neutral
bodies endowed with magnetic field. Starting from the similarity of neutron
particles with neutron stars for which the general-relativistic framework is
imposed, this treatment is extended to neutrons. A class of interior
solutions is derived from the Einstein field equations for a
spherically-symmetric distribution of a perfect magneto-fluid in the
magnetohydrodinamic approximation and a magnetic mass model is proposed. The
dependence of the metric and of physical parameters on the magnetic field
leads to the conclusion that the mass is entirely magnetic in origin, which
implies that the field equations do not admit interior solutions if the
source consist only on perfect fluid.
\end{abstract}

\section{ Introduction}

The use of general relativity framework for the study of particles is an
attractive idea and is of considerable historical interest in connection
with efforts to built models of charged particles which derive their mass
from electromagnetism.

Let us remind the problem of the structure of electron whose mass many
scientists believe is of purely electromagnetic origin. Various attempts to
build charged sphere electromagnetic mass models for the electron within the
realm of classical electromagnetism were not fruitful, since the evenly
charged parts of a charged sphere repel one another, due to the Coulomb
repulsive forces. Poincare showed that the introduction of
non-electromagnetic coercive forces ( a kind of negative pressure) was
necessary to provide stability and to build a consistent picture of such a
model. Thus, the Einstein field equations should be considered for $\rho
\left( 0\right) =0$ and nonvanishing stress which acts in conjunction with
gravity to counteract the Coulomb repulsion and maintain equilibrium.

The tentative of embedding particles in the gravitational theory was already
successfully fructified by a number of famous theoreticians. Schwarzschild
found the interior metric for the Schwarzschild solution representing a
sphere with a constant non-gravitational energy density throughout.
Cooperstoch and De La Cruz \cite{1}, generalizing this solution to the case
of a charged sphere of dust, found not only physically motivated static
sources for the Reissner - N\"{o}rdstrom metric, but also a family of
solutions for which the mass is electromagnetic in origin and the pressure $p
$ is negative in interior, increasing monotonically outward until $p=0$ at
the surface. Tiwari, Rao and Kanakamedala \cite{2} elaborated an
electromagnetic mass model which could be extended to a class of charged
spherical particles whose mass is entirely electromagnetic in origin and the
Coulomb repulsive forces are balanced by an in-built negative pressure of
the charged fluid distribution. Recently, Kauffmann \cite{3} studied the
self-gravitational effects of the electron electromagnetic field, that acts
as an electromagnetic mass contribution. Campanelli and Lousto \cite{4}
looked for non-perturbative interior solutions and found the approximate
metric in the case of static, uniform density models of neutron stars.

The intention of the present paper is to take advantage of the similarity
between stellar bodies and particles, especially when this is evident, which
is the case of the pair neutron particle and neutron star \cite{5}. The
general approach which is developed here consist on analyzing the interior
solutions of Einstein field equations with sources, using those
energy-momentum tensors which describe the studied objects, with the
intention to justify the existence of a magnetic mass model for neutron
particle and neutron star \cite{6}.

\section{General considerations}

The problem of finding interior solutions of Einstein field equations is of
great interest. One of the most useful analytic solution is the
Schwarzschild metric which describes a star of uniform density, $\rho =\rho
_o=const$ $for$ $all$ $p$, with the Reissner-Nordstrom generalization.

The unique Schwarzschild and Reissner-N\"{o}rdstrom solution for the empty
space Einstein and Einstein - Maxwell field equations respectively, can be
written in the Schwarzschild coordinates, as (the signature is $+2$ and the
constants are normalized to unity):

\begin{equation}
ds^2=g_{ab}dx^adx^b=\frac 1{f\left( r\right) }dr^2+r^2\left( d\theta ^2+\sin
^2\theta d\varphi ^2\right) -f\left( r\right) dt^2  \label{eq}
\end{equation}

\[
a,b=0,1,2,3 
\]
where: 
\begin{equation}
f\left( r\right) =1-\frac{2m}r  \tag{1.a}  \label{eq}
\end{equation}
in the case of the Schwarzschild solution and:

\begin{equation}
f\left( r\right) =1-\frac{2m}r+\frac{q^2}{r^2}  \tag{1.b}  \label{eq}
\end{equation}
in the case of the Reissner-N\"{o}rdstrom solution. Here $m$ is the mass of
the source and $q$ is the total charge residing on it.

This form of the metric is quite simple and interesting to deal with so it
appear natural to suppose that the matter distribution of the source which
generates the exterior fields is also described by a metric of the same form:

\begin{equation}
ds^2=\frac 1gdr^2+r^2\left( d\theta ^2+\sin ^2\theta d\varphi ^2\right)
-gdt^2  \label{eq}
\end{equation}
where $g$ is an unknown function. Starting from this metric, Tiwari \cite{2}
finded a class of charged spherical particles which obey to the following
two conditions: (1) if the source consist of a perfect fluid distribution,
there exists no solution to the Einstein matter field equations; and (2) if
the source consist of a charged perfect fluid distribution, a class of
solutions to the Einstein-Maxwell field equations exists werein all the
physical parameters (mass, density, pressure, etc.) are charge dependent and
vanish when the charge vanishes.

In this paper the purpose is to demonstrate the existence of a magnetic-mass
model, which leads to similar interesting consequences as presented above.

The source is considered a perfect, cuasi-neutral fluid, endowed with
magnetic field. This kind of assumption could lead to the characteristics of
plasma fluid \cite{7}, which is a perfect conductor medium. Consequently, in
the magnetohydrodynamic approximation, the electric field become identic
null and the electromagnetic tensor contain only magnetic terms \cite{8}.

The energy-momentum tensor that describes the behavior of a perfect
magneto-fluid is the corresponding perfect fluid energy-momentum tensor with
the nonvanishing components of the electromagnetic tensor.

\begin{equation}
T_{ab}=\left( \rho +\frac{B^2}{8\pi }\right) u_au_b+\left( p+\frac{B^2}{8\pi 
}\right) h_{ab}-\frac{B_aB_b}{4\pi }  \label{eq}
\end{equation}
where $\rho $ is the energy density, $p$ is the pressure, $B_a$ is the
magnetic field 4-vector, $u_a$ is the velocity 4-vector field tangent to the
fundamental flow-lines, which is normalized $u_au^a=-1$ and $h_{ab}$ is the
projection tensor into the tangent three-space orthogonal to $u^a$:

\begin{equation}
h_{ab}=g_{ab}+u_au_b  \label{eq}
\end{equation}

The space-time generated by this source can be considerated stationary and
axisymmetric, which means that there are two Killing vector fields
associated with the coordinates $\left( t,\varphi \right) $ \cite{9}. For
this reason, the components of the metric tensor and consequently, the
physical quantities, will depend only on coordinates $\left( r,\theta
\right) $. Two particular situations will be considered: the neutron
particle and the neutron star case and a magnetic mass model will be
developed for both objects.

\section{The neutron particle case}

First will be analyzed the particular case when the quantities $g,\rho $ , $p
$ and $B$ depends only on $r$ (all the calculations have been made within
the GRTensor package \cite{10}).

The nonvanishing components of the Einstein tensor for the above metric are:

\begin{eqnarray}
G_r^r &=&G_t^t=\frac 1{r^2}\left( rg^{\prime }(r)+g(r)-1\right)  \label{eq}
\\
G_\theta ^\theta &=&G_\varphi ^\varphi =\frac 1{2r}\left( 2g(r)^{\prime
}+rg(r)^{\prime \prime }\right)  \label{eq}
\end{eqnarray}

To keep the energy-momentum tensor diagonal, $B_\theta $ will be taken also
identic null and the only nonvanishing component for the magnetic field will
be $B_r=B\left( r\right) $. The energy-momentum tensor equations are:

\begin{eqnarray}
T_r^r &=&\frac 1{8\pi }\left( 8\pi p\left( r\right) -g\left( r\right)
B\left( r\right) ^2\right)  \label{eq} \\
T_\theta ^\theta &=&T_\varphi ^\varphi =\frac 1{8\pi }\left( 8\pi p\left(
r\right) +g\left( r\right) B\left( r\right) ^2\right)  \label{eq} \\
T_t^t &=&-\frac 1{8\pi }\left[ 8\pi \rho \left( r\right) +B\left( r\right)
^2g\left( r\right) \right]  \label{eq}
\end{eqnarray}

From the conservation of the energy momentum tensor equation $T_{b;a}^a=0,$
only the radial component is relevant, the others been identic null:

\begin{equation}
p\left( r\right) ^{\prime }-\frac{g\left( r\right) B\left( r\right) }{4\pi }%
\left( B\left( r\right) ^{\prime }-\frac 2rB\left( r\right) \right) -\frac{%
g^{\prime }\left( r\right) }{8\pi }\left( B\left( r\right) ^2-\frac{4\pi
\left( p\left( r\right) +\rho \left( r\right) \right) }{g\left( r\right) }%
\right) =0  \label{eq}
\end{equation}

When $p=0$, if the first Einstein equation $G_r^r=8\pi T_r^r$ is substituted
in the energy-momentum tensor conservation equation, on obtain athe metric
function at the radius $R$ and magnetic field $B_s$ that corresponds to the
surface of the particle sphere.

\begin{equation}
g\left( R\right) =\frac 1{2a}\left( -b\pm \sqrt{b^2-4ac}\right)  \label{eq}
\end{equation}
where: $a=2RB_sB_s^{\prime }+3B_s^2-B_s^4R^2$ , $b=B_s^2+4\pi \rho \left(
1+B_s^2R^2\right) $ and $c=4\pi \rho .$ The identification with the exterior
neutral Schwarzchild solution imposes the condition of vanishing surface
magnetic field: $B_s=0$.

The difference $G_r^r-G_t^t=8\pi G\left( T_r^r-T_t^t\right) =0$ shows that
in interior the pressure is negative: $p\left( r\right) =-\rho \left(
r\right) ,$ in accord with the results of Cooperstock and Tiwari. This is
the case of the neutron particle where the pressure counteract the magnetic
stress $B^2/8\pi $ and should be correlated with the predominant role of
gravitation in maintaining the equilibrium. With this assumption, the
expression for $g^{\prime }(r)$ from the equation $G_r^r=8\pi T_r^r$ is
substituted in the energy momentum tensor conservation equation and a formal
expression for the metric function is obtained:

\begin{equation}
g\left( r\right) =\frac{8\pi r(p^{\prime }(r)-rp(r)B^2(r))-B^2(r)}{%
B(r)\left( 2rB^{\prime }(r)+3B^2(r)-B\left( r\right) ^3r^2\right) }
\label{eq}
\end{equation}

From this equation we observe that at the origin, where $\rho =0$\textit{, }%
the expression of the metric will depend only on the magnetic field: $%
g_0=1/3B$\textit{, }thus the magnetic field is essential for the stability
of the neutral particle.

Therefore, if the magnetic field became identic null and the source consist
only on a perfect fluid distribution, there exist no metric solutions: $%
B(r)\rightarrow 0$ $\Rightarrow g\left( r\right) \rightarrow \infty $, which
means that the mass is entirely magnetic in origin.

This dependence of the metric function on $B\left( r\right) $ determines
also a surface singularity: $B(r)=\left( r^3+r^2+c\right) ^{-\frac 12}$,
which stands the magnetic field role on the limitation of the particle
radius. In fact, if the pressure takes the particular value: $p(r)=e^{\int
B^2(r)rdr}+const$ , its influence will be elimiated and the metric will
depend only on $B\left( r\right) $: $g\left( r\right) =-\left( 2r\frac{%
B^{\prime }(r)}{B\left( r\right) }+3B\left( r\right) -B\left( r\right)
^3r^2\right) ^{-1}$, which shows the essential contribution of the magnetic
field to the geometry.

If this expression of the metric function is replaced in the Einstein field
equations and yields a dependence of the physical parameters (density and
pressure) on the magnetic field. The explicit dependence of $p\left(
r\right) $ and $\rho \left( r\right) $ on $B\left( r\right) $ is too
complicated to be presented but is simply to verify that even those
parameters do not admit solution when the magnetic field vanishes.

From the second Einstein field equation $G_\theta ^\theta =8\pi T_\theta
^\theta $, a Poisson-type equation is obtained:

\begin{equation}
\nabla ^2g\left( r\right) =16p\left( r\right) \pi +2B\left( r\right) ^2g(r)
\label{eq}
\end{equation}

This equation clearly shows the sources of the gravitational field and
states the role of the magnetic field as a magnetic mass.

\section{The neutron star case}

With the discovery of compact astrtophysical objects like binary pulsars,
white dwarfs and black holes, finding exact interior non-perturbative
solutions of the Einstein field equations \cite{11} become a necessarily
subject for astrophysicists and relativists. To study the interior field of
such a strange object it is difficult, because we don't know if in so
extreme conditions of density and pressure the conservation laws for the
energy-momentum tensor are valid. We don't even know the state equation, but
we'll use the general ansatz: $p=\nu \rho $ where $\nu $ is a constant%
\textit{. }As a first approximation we will suppose that the metric is
static and axially symmetric and the source is a magnetized perfect fluid.
This means that the space-time contains one space-like and one time-like
commuting hypersurface forming two Killing vectors so in this case the
dependencies of $g,\rho $, $p$ and $B_a$ on the remaining $\left( r,\theta
\right) $ coordinates will be considered. The Einstein tensor has a more
complicated expression, with nonvanishing non-diagonal components. The
electro-magnetic four potential is given by: $A_a=(0,0,A_\varphi ,0)$, which
reads the following form for the magnetic field cuadri-vector: $%
B_a=(B_r,B_\theta ,0,0)$. It is suitable for further calculations, to
separate the variables and to split the metric function in two: $g\left(
r,\theta \right) =g_r\left( r\right) g_\theta \left( \theta \right) $

The difference $(G_\theta ^\theta -G_\varphi ^\theta )=8\pi (T_\theta
^\theta -T_\varphi ^\theta )$ is a simple expression: 
\begin{equation}
\dot{g}\left( r,\theta \right) =2g\left( r,\theta \right) B_\theta \left(
r,\theta \right)  \label{eq}
\end{equation}
which is integrated to give the dependence of metric function on $\theta $:

\begin{equation}
g_\theta \left( \theta \right) =e^{2\int_0^{2\pi }B_\theta \left( r,\theta
\right) d\theta }  \label{eq}
\end{equation}

To deduce the dependence of $g\left( r,\theta \right) $ on variable $r$, it
is useful to consider the non-diagonal Einstein field equation: $E_\theta
^r=8\pi T_\theta ^r$:

\begin{equation}
\dot{g}^{\prime }\left( r,\theta \right) =4g\left( r,\theta \right)
B_r\left( r,\theta \right) B_\theta \left( r,\theta \right)  \label{eq}
\end{equation}
and the integration of this equation yields:

\begin{equation}
g_r\left( r\right) =\frac 1{B_\theta \left( r,\theta \right)
}e^{2\int_0^RB_r\left( r,\theta \right) dr}  \label{eq}
\end{equation}

The solution is therefore:

\begin{equation}
g\left( r,\theta \right) =\frac 1{B_\theta \left( r,\theta \right)
}e^{2\left( \int_0^RB_r\left( r,\theta \right) dr+\int_0^{2\pi }B_\theta
\left( r,\theta \right) d\theta \right) }  \label{eq}
\end{equation}
which can be integrated for particular values of the magnetic field.

This result can be interpreted as follows: if the source consist on a
magneto-fluid distribution, a class of solutions to the Einstein field
equation exist wherein the metric and the physical parameters are all
dependent on the magnetic field and there is no solution when the azimuthal
component of the magnetic field vanishes. The existence of the azimuhal
magnetic field is a realistic condition for the neutron star, this being the
only stable configuration for a stationary axisymmetric flow of the internal
stellar plasma in the MHD approximation.

The Poisson-type equation has a radial and an azimuthal component: $\nabla
^2g\left( r,\theta \right) =\nabla _r^2g\left( r,\theta \right) +\nabla
_\theta ^2g\left( r,\theta \right) $. The radial one, obtained from the sum: 
$G_\theta ^\theta +G_\varphi ^\varphi =8\pi \left( T_\theta ^\theta
+T_\varphi ^\varphi \right) $ is similar to the first case: 
\begin{equation}
\nabla _r^2g\left( r,\theta \right) =16p\left( r,\theta \right) \pi
+2B_r\left( r,\theta \right) ^2g(r,\theta )  \label{eq}
\end{equation}
and the azimuthal Poisson-type equation, obtained from the relation:

\begin{equation}
\left( G_r^r-G_t^t\right) -2\left( G_\theta ^\theta -G_\varphi ^\varphi
\right) =8\pi \left[ (T_r^r-T_t^t)-2\left( T_\theta ^\theta -T_\varphi
^\varphi \right) \right]  \label{eq}
\end{equation}
has the form:

\begin{equation}
\nabla _\theta ^2g\left( r,\theta \right) =8\pi \left( p\left( r,\theta
\right) +\rho \left( r,\theta \right) \right) +\frac 6{r^2}B_\theta ^2\left(
r,\theta \right) ^2  \label{eq}
\end{equation}

Even in this more complicated situation, the same conclusion can be drown
from the Poisson equation: the source of the gravitational field has a
magnetic component which contribute, as a magnetic mass, to the creation of
the geometry.

\section{Conclusions}

This study was made with the intention of analyzing the physical situation
of the neutron particle or the neutron star, both being spherical objects
endowed with significant magnetic field, in order to find a realistic
interior model. The main idea was the use of the general relativity
formalism and the purpose was to obtain a class of static, spherically
symmetric interior solutions. Two situations for the magnetic distribution
were considered: with radial and with both radial and azimuthal components.
The fact that the magnetic field has an active contribution to the metric
and to the physical parameters, was of clear evidence and this dependence
leaded to the conclusion that the mass is entirely magnetic in origin. The
result that if the magnetic field vanishes there is no metric solution, is
important, because shows again the fundamental role which the magnetic field
is playing to the constitution of matter and of the Universe.

\end{document}